\documentclass[namedreferences]{SolarPhysics}
\usepackage[optionalrh,solaenum]{spr-sola-addons}
\usepackage{graphicx}                  
\usepackage{amssymb}                    
\usepackage{color}                       
\usepackage{url} % For breaking URLs easily trough lines
 
\begin{document}
\begin{article}
\begin{opening}
\title{Hysteresis in a Solar Activity Cycle}
\author{Vinita~\surname{Suyal}$^1$\sep
        Awadhesh~\surname{Prasad}$^2$\sep
        Harinder~P.~\surname{Singh}$^3$
       }
\runningauthor{Suyal et al.}
\runningtitle{Hysteresis in a Solar Activity Cycle}
\institute{$^{1,2,3}$Department  of Physics  and Astrophysics, University of Delhi, Delhi 110007, India.\\
email:$^1$vsuyal@physics.du.ac.in\\
$^2$awadhesh@physics.du.ac.in\\
$^3$hpsingh@physics.du.ac.in}
%%%%%%%%%%%%%%%%%%%%%%%%%%%%%%%%%%%%%%%%%%%%%%%%%%%%%%%%%%%%%%%%
\begin{abstract}
We analyze \textit{in situ} measurements of solar wind velocity obtained by
the \textit{Advanced Composition Explorer} (ACE) spacecraft during the
solar activity cycle $23$. We calculated a robust complexity
measure, the permutation entropy ($S$)
of solar wind time series at different phases of a solar activity
cycle. The permutation entropy measure is first tested on the known dynamical data before its
application to solar wind time series.
It is observed that complexity of solar wind velocity fluctuations
at $1$ AU shows hysteresis phenomenon while following
the ascending and descending 
phases of the activity cycle. This indicates the presence of
multistability in the dynamics governing the solar wind
velocity over a solar activity cycle.

\end{abstract}
\end{opening}
%%%%%%%%%%%%%%%%%%%%%%%%%%%%%%%%%%%%%%%%%%%%%%%
\section{Introduction}
\label{S-intro}
The corona changes its shape enormously during a solar activity 
cycle resulting in temporal and structural properties of the
solar wind velocity variation in a solar cycle \cite{schw07}. 
\textit{In situ} solar wind plasma observations show that its local 
properties at $1$~AU are modulated by the solar activity
cycle \cite{hapg91,rich08}.
Around solar activity minimum, the structures of the
corona and the solar wind are rather simple and remain
so for several months \cite{schw07}.
At solar activity maximum, slow solar wind dominates at all
helio latitudes \cite{mcco00}. Recently it was reported
that slow solar wind velocity just before the maximum of the solar
activity cycle is least correlated to data obtained 
from the rest of the solar activity cycle \cite{suya11}.

Observations of solar wind velocity made by
different spacecraft have been analyzed in considerable detail
and reported by several authors.
Values of complexity measures such as entropy
(\opencite{mace97}, \citeyear{maco98}; \opencite{mace00}; \opencite{reda01}), correlation dimension
\cite{mace97,gupt08}
and Lyapunov exponents \cite{mace98,reda01,gupt08}
show that solar wind velocity fluctuations are a consequence of complex
nonlinear dynamical processes.
Inherent changes in the dynamics governing the solar wind velocity
at $0.3$ AU have been observed \cite{gupt08}.
\inlinecite{mila04} analyzed magnetic
and bulk velocities measured by ACE and found
an anisotropy in the velocity, magnetic, and cross helicity
correlation functions and power spectra.
\inlinecite{mcco00} analyzed Ulysses observations to demonstrate
that the mid-latitude solar wind structure becomes increasingly
complex as solar activity increases. \inlinecite{cons09} investigated
the emergence of spatio-temporal complexity in the $11$-year solar
cycle monitored by sunspot activity. They showed that spatio-temporal
or dynamical complexity is an intrinsic property of the solar cycle.
Using information entropy approach to the sunspot number time series,
they showed how the dynamical complexity increases
during the maximum phase of the solar cycle.

Hysteresis occurs in several phenomena in physics, chemistry,
biology, and engineering.
It is a nonlinear phenomenon observed in systems
from diverse areas of science,
\textit{e.g}., electromagnetism, electro-plasticity, superconductivity,
and granular motion \cite{bert98,guye99,katz02,zhar02}.
This phenomenon occurs when a nonlinear
system has at least two coexisting stable states in the hysteresis region where
the system is found to depend on the history of the dynamics. Here, in dynamical 
systems, history corresponds to the  system's initial conditions.
For example, hysteresis is observed in van der Pol
system, Duffing system \cite{thom86}, and Lorenz system
\cite{alfs85}. The hysteresis has been
observed in coupled nonlinear systems as well  \cite{pras05}.

Hysteresis phenomenon has been noticed in various solar indices.
\inlinecite{bach94} observed the presence of hysteresis patterns
among many pairs of activity indices during solar cycle $21$ and $22$.
They found that this hysteresis can be expressed
approximately as a hierarchy
of delay times behind the leading index, the sunspot number.
\inlinecite{reye98} analyzed the low-degree $p$-mode frequency shifts 
and solar activity indices (radio flux at $10.7$ cm and magnetic index)
over solar cycle $22$ and observed a hysteresis phenomenon. \inlinecite{more00}
suggested that high latitude fields are necessary to produce a significant difference
in hysteresis between odd and even-degree $p$-modes frequencies.
\inlinecite{trip00} reported that the intermediate degree $p$-mode
frequencies of solar cycle $22$ show a hysteresis phenomenon with the
magnetic indices whereas no such effect exists for the radiative indices.
\inlinecite{ozgu01} showed the presence of hysteresis between the solar
flare index and some solar activity indicators such as total sunspot area,
mean magnetic field, and coronal index during solar cycles $21$ and $22$.
They found that these indices follow different paths for ascending and descending
phases of the solar cycles while saturation effect exists at the extreme phases.

In the present work, we attempt to understand the dynamics
of ascending and descending phases of a solar cycle.
We use permutation entropy $(S)$ of hourly
averaged solar wind velocity time series
to capture the complexity trend over a cycle.
We use the data obtained from ACE during $1998-2010$.
This period belongs to solar activity cycle $23$.
We use permutation entropy $(S)$
to detect the hysteresis in a dynamical system and calculate
it for time series obtained
from simulated as well as solar wind velocity data.

In the next section, we review the algorithm
to calculate the permutation entropy $(S)$ of a time series.
In Section~\ref{S-dynamical_hysteresis} we describe how 
$S$ detects the multistability present in the modeled
dynamical system. In Section~\ref{S-solar_wind} we analyze
solar wind data using permutation entropy.
This is followed by  conclusions in Section~\ref{S-conc}.
%%%%%%%%%%%%%%%%%%%%%%%%%%%%%%%%%%%%%%%%%%%%%%%%%%%%%%%%%%%%%%%%%
\section{Permutation Entropy}
\label{S-permutation_entropy}
Lyapunov exponent, entropy, and fractal dimension are
well known and extensively used measures to detect
dynamical changes in a time series obtained from
a complex system \cite{kant04}. Calculation of these
quantities needs phase space reconstruction for which we need to know the
parameters like embedding dimension and delay \cite{pack80}.
However, in practice, it is very difficult
to get  accurate parameters, particularly for noisy data.
Permutation entropy can be used
to compare two or more time series and distinguish regular, chaotic and random
behavior for small and noisy time series \cite{band02}.
It quantifies not the only randomness but also
the degree of correlational structures of a
time series \cite{ross07}.
Permutation entropy is conceptually simple and computationally very fast \cite{cao04}.
It can be effectively used to detect dynamical
changes in a complex system.
The detailed algorithm to calculate permutation entropy is described below :

Let us consider a time series  ${x_i, i=1,2,...,N}$ and embed it in an
$m$-dimension\-al space \cite{pack80}. An embedded vector is written as
$$X_i=[x_i,x_{i+1},...,x_{i+(m-1)}].$$
For each embedded vector $X_i$, the $m$ components
can be arranged in an increasing order:
$[x_{i+k_1}< x_{i+k_2}<...< x_{i+k_m}]$,
where $k_j$ can have any value from $0$ to $m-1$.
Hence, each vector $X_i$ is uniquely mapped onto $(k_1,k_2,...,k_m)$ 
which is one of the $m!$ permutations of $m$ distinct symbols
$(0,1,...,m-1)$. When each permutation is considered as a symbol, the 
reconstructed trajectory in the $m$-dimensional space is
represented by a symbol sequence. We denote the probability
distribution for the distinct symbol 
by $P_1,P_2,...,P_j$, where $j\leq m!$.
The permutation entropy for the time series ${x_i, i=1,2,...,N}$
is defined \cite{band02}
as the Shannon entropy for the
$j$ distinct symbols
\begin{equation}
S(m)=- \sum_{j=1}^{m!}{P_j\mbox{ln}(P_j)}.
\end{equation}
Here $S(m)$ attains the maximum value $\mbox{ln}(m!)$ when $P_j=\frac{1}{m!} \forall j$;
%\textit{\textit{i.e}}., 
for uniformly distributed data.
Therefore, normalized permutation entropy is written as
\begin{equation}
S=\frac{S(m)}{\mbox{ln}(m!)},
\end{equation}
where the value of $S$ lies between $0$ and $1$.
Smaller value of $S$ indicates a more regular time series.
If $m$ is too small there are very few distinct states
and this scheme will not work.
A value of $m=5,6,$ or $7$ is suitable to calculate 
the permutation entropy for detecting
the dynamical changes in a system \cite{band02,cao04}.
%%%%%%%%%%%%%%%%%%%%%%%%%%%%%%%%%%%%%%%%%%%%%%%%%%%%%%%%%%%%%%%%
\section{Dynamical Hysteresis}
\label{S-dynamical_hysteresis}
Hysteresis occurs when a nonlinear system has at least two existing
stable states in a hysteresis region.
Out of multiple states, the system attains a
state depending on the history
of the dynamics, \textit{e.g}., on the initial conditions of the system
\cite{thom86,pras05}.
We consider one of the example from  \inlinecite{pras05}, where  the dynamical hysteresis in
coupled oscillators have been observed in terms of
stability of the system using Lyapunov exponents
in a wide range of parameter space. It is conceptually difficult
to estimate the Lyapunov exponent for a small and noisy data set.
In order to see the hysteresis region
using the permutation entropy, we consider  R\"ossler-type \cite{gasp83}
coupled oscillators, as shown schematically in Figure 1.
The model equations for this system are 
\begin{eqnarray}
\nonumber
\frac{dx_i(t)}{dt} &=& -w_iy_i-z_i+F_i(\epsilon,x_i,x_j),\\
\nonumber
\frac{dy_i(t)}{dt} &=& w_ix_i+a_iy_i,\\
\frac{dz_i(t)}{dt} &=& \beta_ix_i+z_i(x_i-y_i),
\label{oscillator}
\end{eqnarray}
where  $i,j$=1,2,3,4 (number of oscillators$=4$),
$w_1 = 1.005, w_2 = w_3 = w_4 = 0.995, \alpha_i =0.38,
\beta_i =0.3,$ and $\gamma_i =4.5.$
At these set of parameters all the individual systems show 
chaotic oscillations \cite{pras05}.
The coupling functions are $F_1 = \epsilon(x_2+x_3+x_4-3x_1),
F_2=\epsilon(x_1-x_2), F_3=\epsilon(x_1-x-3), $
and $F_4=\epsilon(x_1-x_4)$ while  $\epsilon$ is the coupling parameter.
In our numerical calculation, we used the fourth-order Runge-Kutta integrator
with integration time step $0.01$.
\begin{figure}
\vspace{.4cm}
\centering
\includegraphics[width=5cm,height=2.7cm]{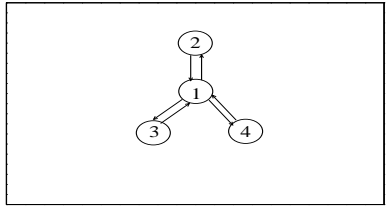}
\caption{Schematic configuration of coupled oscillators represented by
Equation (\ref{oscillator}).
}
\label{fig1}
\end{figure}
\begin{figure}
\vspace{.8cm}
\centering
\includegraphics[width=12cm,height=6cm]{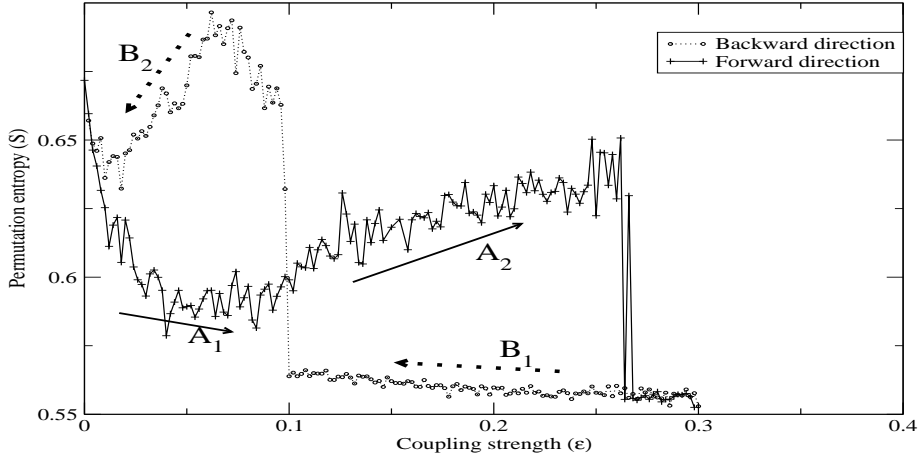}
\caption{Permutation entropy ($S$) for increasing (solid line)
and decreasing (dotted line) coupling strengths
($\epsilon$) for the coupled system represented by
Equation (\ref{oscillator}). Arrows with symbols A and B
indicate the paths for  increasing and decreasing coupling strengths, respectively.}
\label{fig2}
\end{figure}
\begin{figure*}
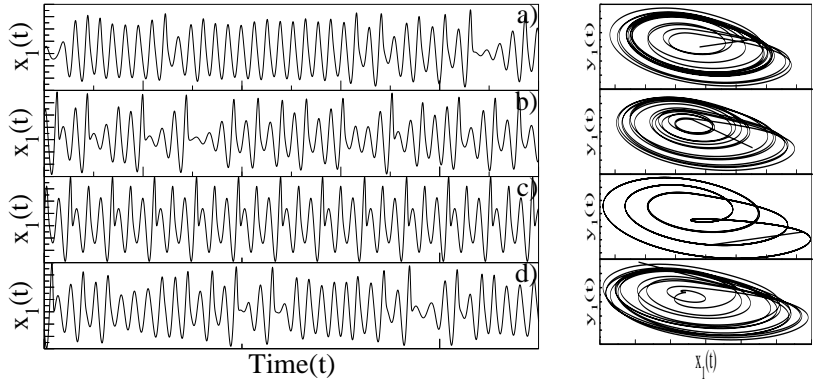

\vspace{20pt}
\begin{minipage}{.65\linewidth}
\centering
\includegraphics[width=7cm,height=5cm]{3_a}
\end{minipage}
\begin{minipage}{.3\linewidth}
\includegraphics[width=3cm,height=5cm]{3_b} 
\end{minipage}
\caption{The time series
$x_1(t)$ in regions (a) $A_1 (\epsilon =0.05)$, (b) $A_2 (\epsilon =0.20)$, (c)
$B_2 (\epsilon =0.20)$, and (d) $B_1 (\epsilon =0.05)$, respectively of Figure \ref{fig2}. 
The right-hand-side panel shows the corresponding
projection of trajectories into the $x-y$ plane.}
\label{fig3}
\end{figure*}
\inlinecite{pras05} have shown
the hysteresis behavior in the largest Lyapunov exponent with
respect to the coupling parameter $\epsilon$, and predicted the 
coexistence of more than one state of different type of stability in 
the hysteresis region.
In order to check the presence of hysteresis and
dynamical changes in terms of permutation entropy $S$, we
consider the $x$-component of the time series of the first oscillator. 
We take embedding dimension $m=5$ (although higher values of
$m$ gives similar results)
and sampling time $\tau =300$ (first minima of the
auto-correlation function). $750$ data points are used
to calculate $S$. Figure \ref{fig2} shows a plot of $S$ as 
a function of increasing and decreasing coupling strengths $\epsilon$.
The arrows show the corresponding increasing and decreasing coupling strengths.
The hysteresis region is visible for different values of $\epsilon$.
Figures \ref{fig3}$(a-d)$
show the time series $x_1(t)$
and the projection of the first oscillator in the $x-y$ plane at
regions corresponding to $A_1 (\epsilon=0.05)$, $A_2 (\epsilon=0.2)$,
 $B_2 (\epsilon=0.2)$, and $B_1 (\epsilon=0.05)$ of Figure \ref{fig2}.
Although the trajectories look similar, the corresponding 
permutation entropy ($S$) shows that they are dynamically different.
Importantly, the permutation entropy captures the hysteresis
loop as shown with Lyapunov exponents in \inlinecite{pras05}.
%%%%%%%%%%%%%%%%%%%%%%%%%%%%%%%%%%%%%%%%%%%%%%%%%%%%%%%%%%%%%%%%%%%%%%
\section{Permutation Entropy of the Solar Wind Data}
\label{S-solar_wind}
\begin{figure*}
\centering
\includegraphics[width=9cm,height=5cm]{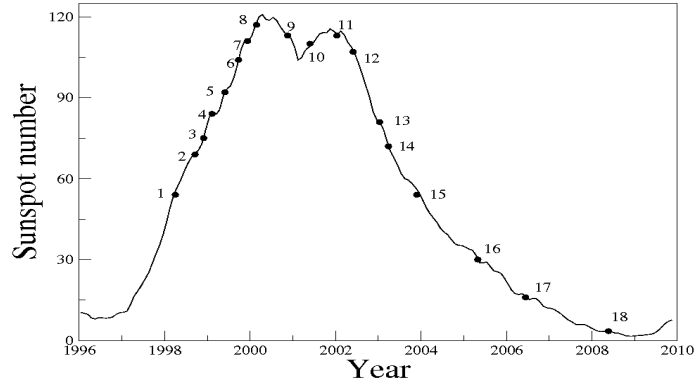}
\caption{Smoothed monthly sunspot number from $1996$ to $2010$.
Serial numbers $1-18$ correspond to the  solar wind data sets given in Table 1.}.
\label{fig4}
\end{figure*}

We use hourly averaged solar wind velocity data at a distance of $1$ AU 
obtained from the Solar Wind Ion Composition Spectrometer
(SWICS) on ACE
(http://www.srl.caltech.edu/ACE/ASC/level2/).
This data set corresponds to the years $1998$ to $2010$.
Since the data are not continuous
we split the data set into $18$ continuous time series
to cover most of the solar cycle $23$.
The details of the time series used are given in Table 1.
In addition, Table 1 (last column)
contains the corresponding sunspot numbers for each
time series taken from http://sidc.oma.be/sunspot-data/.
Figure \ref{fig4} shows the solar activity during activity cycle $23$.
The position of  serial numbers corresponding to the  solar wind data sets,
is shown on the smoothed monthly averaged sunspot index curve.

\begin{table*}[ht]
\caption{Hourly averaged solar wind velocity data measured
by the ACE spacecraft in the years 1998 to 2010:
Initial time ($T_{i}$), Number of Data points ($N$), and Sunspot numbers ($SSN$).
Sunspot numbers are from http://sidc.oma.be/sunspot-data/.}
\resizebox{9cm}{3cm} {
\begin{tabular}{cccc}
\hline
S.No.~~~~~&$T_{i}$~~~~~&$N$~~~~~&Sunspot number (SSN)\\
\hline
~~~~~1~~~~~&1998.25~~~~~&1665~~~~~&54~~~~~\\
~~~~~2~~~~~&1998.71~~~~~&1314~~~~~&69~~~~~\\
~~~~~3~~~~~&1998.91~~~~~&789~~~~~&75~~~~~\\
~~~~~4~~~~~&1999.11~~~~~&1927~~~~~&84~~~~~\\
~~~~~5~~~~~&1999.41~~~~~&2716~~~~~&92~~~~~\\
~~~~~6~~~~~&1999.73~~~~~&1139~~~~~&104~~~~~\\
~~~~~7~~~~~&1999.94~~~~~&1404~~~~~&111~~~~~\\
~~~~~8~~~~~&2000.15~~~~~&3075~~~~~&117~~~~~\\
~~~~~9~~~~~&2000.88~~~~~&1580~~~~~&113~~~~~\\
~~~~~10~~~~~&2001.40~~~~~&1315~~~~~&110~~~~~\\
~~~~~11~~~~~&2002.03~~~~~&3329~~~~~&113~~~~~\\
~~~~~12~~~~~&2002.41~~~~~&3417~~~~~&107~~~~~\\
~~~~~13~~~~~&2003.03~~~~~&789~~~~~&81~~~~~\\
~~~~~14~~~~~&2003.24~~~~~&1840~~~~~&72~~~~~\\
~~~~~15~~~~~&2003.90~~~~~&2633~~~~~&54~~~~~\\
~~~~~16~~~~~&2005.33~~~~~&2541~~~~~&30~~~~~\\
~~~~~17~~~~~&2006.45~~~~~&2278~~~~~&16~~~~~\\
~~~~~18~~~~~&2008.39~~~~~&2108~~~~~&3.5~~~~~\\
\hline
\end{tabular}
}
\label{tab}
\end{table*}
\begin{figure*}
\vspace{.8cm}
\centering
\includegraphics[width=12cm,height=6cm]{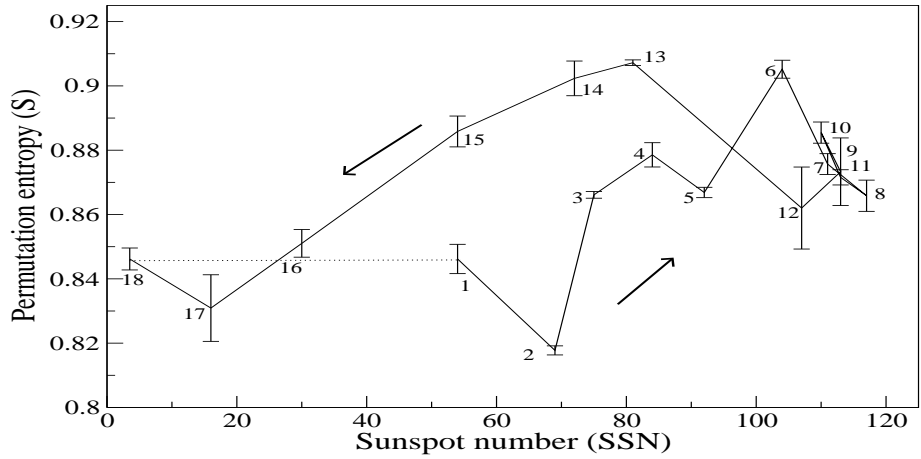}
\caption{Permutation entropy vs. sunspot number for solar wind time series
with $1 \sigma$ error bars. The dotted line is the possible connection between $1$ and $18$.
Arrows show the direction of time.}
\label{fig5}
\end{figure*}
We calculate the normalized permutation entropy $S$ for
all $18$ time series using embedding dimension
$m=5$ and window length of $750$ points.
We use six overlapping windows and
then calculated the average value of $S$. The standard deviation in $S$
is given by
\begin{equation}
 \sigma = \sqrt{\frac{\sum_{i=1}^{n}(S_i-S_{\mbox{avg}})^2}{n}},
\end{equation}
where $n$ is the number of overlapping windows,
$S_{\mbox{avg}}$ is the average permutation entropy of $n$ windows,
and $\sigma$ estimates the statistical uncertainties in the average.
Figure \ref{fig5} shows the plot of $S_{\mbox{avg}}$
of solar wind time series vs. sunspot number corresponding 
to the time of solar wind observation. The error bars
are estimates of the statistical uncertainties in these averages.

It is noticeable that the path followed
by the permutation entropy of the solar wind data in the increasing
phase of the solar activity
cycle is different from that of the decreasing phase, \textit{i.e}., a
hysteresis phenomenon is present.
It shows that although sunspot data set is 
almost symmetric around the peak of the solar activity cycle
yet permutation entropy of solar wind data follows different paths
in increasing and decreasing phases of the solar activity cycle.
%%%%%%%%%%%%%%%%%%%%%%%%%%%%%%%%%%%%%%%%%%%%%%%%%%%%%%%%%%%%%%%%%%%%%%%%
\section{Conclusions}
The long term sunspot time series shows an
average cycle length of $11$ years.
Although nearly periodic, the period as well as amplitude
of the cycle varies irregularly.
Apart from the sunspot data, the intrinsic irregularity of the solar
cycle is seen in other observable variables like
surface flows, solar irradiance (solar constant), the solar wind, and so on.
The Maunder Minimum and several ancient periods
from solar proxy-data suggest
that the Sun exhibits quasi-periodic or intermittent behavior \cite{femi97}.
Owing to changes in magnetic activity,
many aspects of the solar wind
change over a solar cycle, including
the speed, the density, the dynamic pressure, the composition,
and the temperature \cite{rich08}.

In this paper, we present the analysis of solar wind velocity data during the solar
activity cycle $23$. We obtained $18$ different time series of hourly
averaged solar wind velocity, measured by ACE spacecraft at $1$ A.U.
To quantify the randomness of these time series,
we use a robust, conceptually simple and computationally efficient
measure called permutation entropy.
A smaller value of permutation
entropy indicates a more regular time series.
We observe that as the solar cycle $23$ progresses
towards maximum, the permutation entropy increases,
saturates around the peak of activity and then
decreases as the activity of cycle $23$ subsides.
We also note (Figure. \ref{fig5}) that the value of permutation
entropy follows different paths in the ascending and descending
phases of the solar activity cycle.
In addition, while the ascent is fluctuating,
the descent is smooth. This hysteresis phenomenon shows
the multistability in the dynamics of solar wind,
over the solar activity cycle.
The behavior is similar to the one observed for
hysteresis phenomenon of other solar indices
\cite{bach94,reye98,ozgu01} and confirms (\textit{cf}. \opencite{cons09})
that spatio-temporal
or dynamical complexity is an intrinsic property of the solar cycle.

\label{S-conc}
%%%%%%%%%%%%%%%%%%%%%%%%%%%%%%%%%%%%%%%%%%%%%%%%%%%%%%%%%%%%%%%%%%%%
%% Acknowledgements
\begin{acks}
The authors thank the ACE Science Center and instrument teams 
for making available the ACE data used here. VS and AP
thank CSIR for SRF and DST
Govt. of India for financial supports respectively.
\end{acks}

\end{article} 
\end{document}